\documentclass{sf2a-conf2018}
\usepackage{graphicx}
\usepackage{subcaption}
\usepackage{mwe}
\usepackage{hyperref}
\usepackage[]{natbib}  
\usepackage{epstopdf}

\def\BibTeX{{\rm B\kern-.05em{\sc i\kern-.025em b}\kern-.08em
    T\kern-.1667em\lower.7ex\hbox{E}\kern-.125emX}}
\bibpunct{(}{)}{;}{a}{}{,}  

\begin{document}

\TitreGlobal{SF2A 2018}


\title{Probing the structure and evolution of active galactic nuclei with the ultraviolet polarimeter POLLUX aboard LUVOIR}

\runningtitle{AGN science with POLLUX}

\author{F. Marin}\address{Universit\'e de Strasbourg, CNRS, Observatoire Astronomique de Strasbourg, UMR 7550, F-67000 Strasbourg, France}
\author{S. Charlot}\address{Sorbonne Universit\'es, UPMC-CNRS, UMR7095, Institut d'Astrophysique de Paris, F-75014, Paris, France} 
\author{D. Hutsem\'ekers}\address{Space Sciences, Technologies and Astrophysics Research (STAR) Institute, Universit\'e de Li\`ege,All\'ee du 6 Ao\^ut 19c, B5c, 4000 Li\`ege, Belgium}
\author{B. Ag\'is Gonz\'alez$^{3,}$}\address{Instituut voor Sterrenkunde, KU Leuven, Celestijnenlaan 200D, bus 2401 3001 Leuven, Belgium} 
\author{D. Sluse$^3$}
\author{A. Labiano}\address{Centro de Astrobiolog\'ia (CAB, CSIC-INTA), ESAC Campus, E-28692 Villanueva de la Ca\~nada, Madrid, Spain} 
\author{L. Grosset}\address{LESIA, Paris Observatory, PSL University, CNRS, Sorbonne Universit\'e, Univ. Paris Diderot, Sorbonne Paris Cit\'e, 5 place Jules Janssen, 92195 Meudon, France} 
\author{C. Neiner$^6$}
\author{J.-C. Bouret}\address{Aix Marseille Univ, CNRS, CNES, LAM, Marseille, France}

\setcounter{page}{237}


\maketitle


\begin{abstract}
The ultraviolet (UV) polarization spectrum of nearby active galactic nuclei (AGN) is poorly known. 
The Wisconsin Ultraviolet Photo-Polarimeter Experiment and a handful of instruments on board 
the Hubble Space Telescope were able to probe the near- and mid-UV polarization of nearby 
AGN, but the far-UV band (from 1200\,\AA\ down to the Lyman limit at 912\,\AA) remains 
completely uncharted. In addition, the linewidth resolution of previous observations 
was at best 1.89~\AA. Such a resolution is not sufficient to probe in detail quantum mechanical
effects, synchrotron and cyclotron processes, scattering by electrons and dust grains, and dichroic 
extinction by asymmetric dust grains. Exploring those physical processes would require a new, 
high-resolution, broadband polarimeter with full ultraviolet-band coverage. In 
this context, we discuss the AGN science case for POLLUX, a high-resolution UV spectropolarimeter,
proposed for the 15-meter primary mirror option of LUVOIR (a multi-wavelength space observatory 
concept being developed by the Goddard Space Flight Center and proposed for the 2020 Decadal Survey
Concept Study).
\end{abstract}



\section{Introduction}
The far and mid-ultraviolet polarization of nearby active galactic nuclei (AGN) is largely uncharted territory. Only 
two missions were equipped with ultraviolet (UV) polarimeters in the past. 

\begin{figure}[ht!]
\centering
  \includegraphics[trim = 0mm 35mm 0mm 35mm, clip, width=8cm]{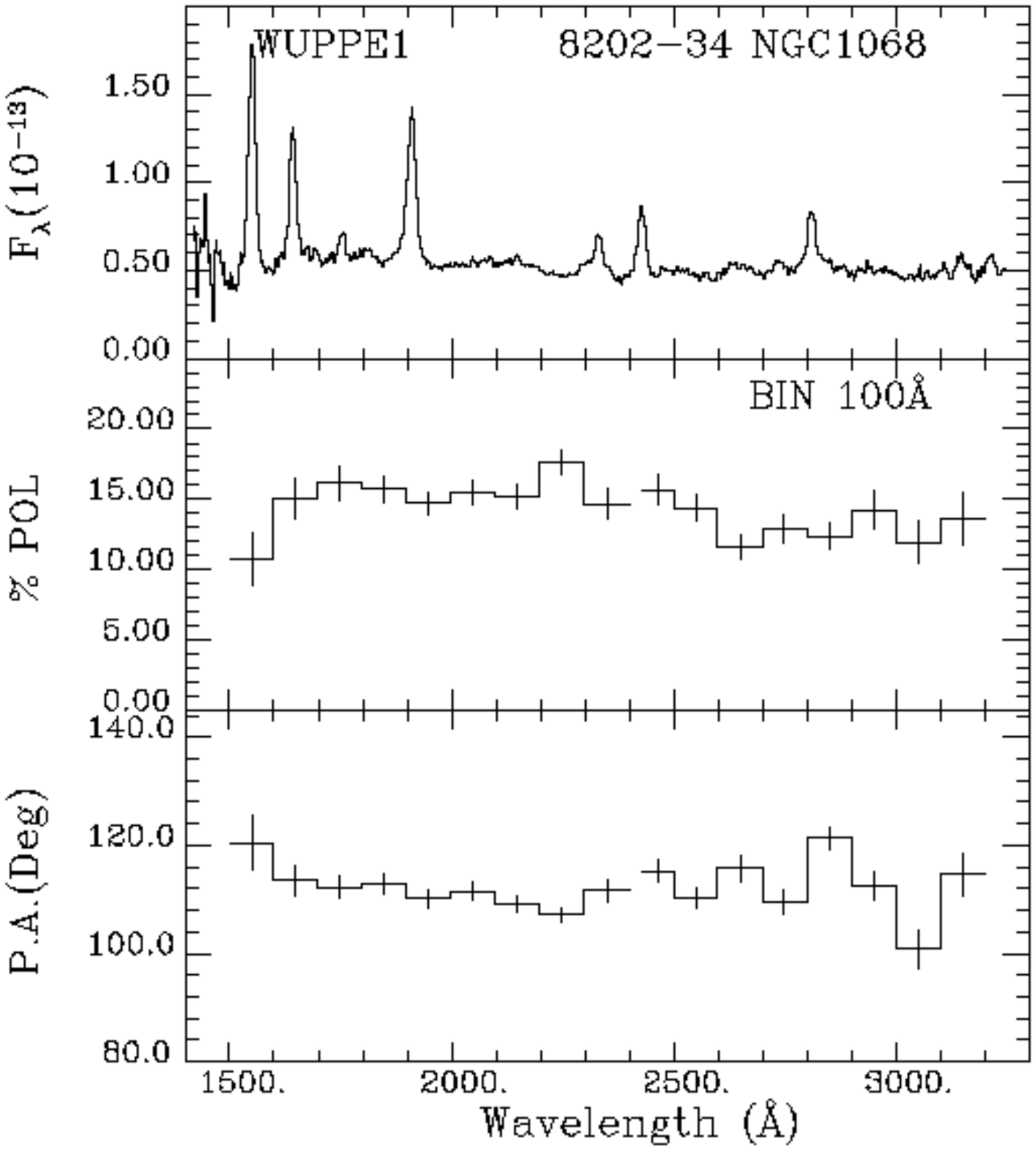}
  \includegraphics[trim = 0mm 35mm 0mm 35mm, clip, width=8cm]{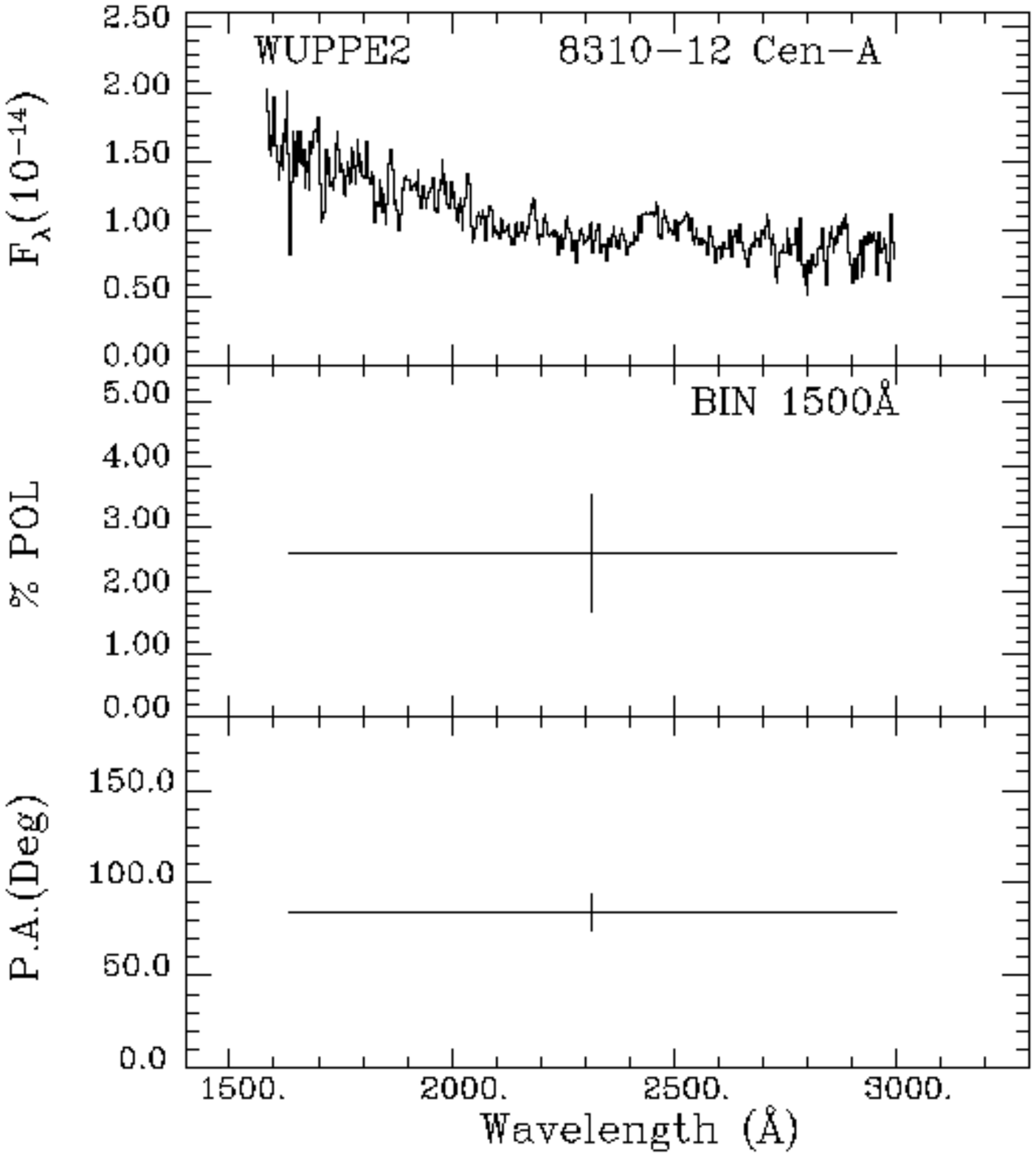}  
  \caption{WUPPE UV spectropolarimetry of the radio-quiet AGN NGC~1068 (left) 
	  and the radio-loud AGN Centaurus~A (right). Both are type-2 AGN
	  (the view of the central engine is blocked by an optically thick,
	  equatorial, dusty medium) and they have comparable GALEX fluxes 
	  (about 28 mJy at 1524~\AA). In the case of NGC~1068, the exposure 
	  time was 1972 seconds. Observation of Centaurus~A was 1152 seconds
	  long. Data from the Barbara A. Mikulski Archive for Space Telescopes
	  (MAST) and from \citet{Code1993}.}
  \label{Fig:WUPPE}
\end{figure}

The first one was WUPPE, the Wisconsin Ultraviolet Photo-Polarimeter Experiment \citep{Nordsieck1982,Stanford1985,Code1989}. 
The telescope, designed and built at the University of Wisconsin Space Astronomy Laboratory in the 1980's (PI: Arthur D. Code), 
was a pioneering effort to explore polarization and photometry at UV wavelengths. WUPPE was designed to obtain simultaneous 
spectra and polarization measurements from 1400 to 3300~\AA. It consisted of a 0.5m f/10 classical Cassegrain telescope and a spectropolarimeter, 
with a field of view of 3.3 by 4.4 arc-minutes and a resolution of 6~\AA. Its effective area was about 100 cm$^2$ at 2300~\AA. WUPPE
flew on two NASA Space Shuttle missions: ASTRO-1 and ASTRO-2. It was one of three ultraviolet telescopes (with the Hopkins Ultraviolet 
Telescope and the Ultraviolet Imaging Telescope) and one X-ray telescope (the Broad Band X-Ray Telescope) on the ASTRO-1 payload 
which flew on board the Space Shuttle Columbia on December 2 -- 11, 1990. The telescope was re-flown on March 2 -- 18, 1995 on board 
the Space Shuttle Endeavour. In total, WUPPE-1 and WUPPE-2 obtained UV spectropolarimetry (and spectra) for 121 objects 
over 183 observations. These 121 objects include only 2 radio-quiet AGN (NGC~4151, NGC~1068), 2 radio-loud AGN (3C~273,
Centaurus~A), and 1 BL Lac object (Mrk~421). These AGN observations, at the exception of NGC~1068 (shown in Fig.~\ref{Fig:WUPPE}),
had very poor spectral resolution. Most of the UV polarimetric measurements had to be spectrally rebinned because of the combined 
effects of source brightness, WUPPE sensitivity limit, and too short integration times. 

The second mission with UV polarimetric capabilities was the Hubble Space Telescope (HST). Two instruments on board HST allowed 
optical, near- and mid-UV polarimetry: the Faint Object Camera (FOC) and the Faint Object Spectrograph (FOS). Both 
instruments were among the four original axial instruments on board HST and they were designed to take observations from 1150 to 
6500~\AA. The FOS was removed from HST during the Second Servicing Mission in February 1997, and the FOC during Servicing 
Mission 3B in March 2002. Later on, UV/blue filters ($\lambda >$ 2000~\AA) were mounted on the Advanced Camera for Surveys (ACS) 
and the Wide Field and Planetary Cameras  (WFPC) 1 and 2, for polarimetric observations. Altogether,\footnote{Accounting for the
Near Infrared Camera and Multi-Object Spectrometer (NICMOS), which provides broad-band imaging polarimetry in the wavelength range 
0.8 -- 2.5~$\mu$m} the polarimetric instruments on board HST observed 117 AGN (108 objects with imaging-polarimetry, 76 objects
with spectropolarimetry, and a handful with both; Enrique Lopez-Rodriguez, private communication) from Cycle 0 through Cycle 22. 
HST UV polarimetry provided strong constraints on the polarization mechanism in AGN \citep{Antonucci1994}, highlighted the 
three-dimensional structure of the nuclear region of NGC~1068 \citep{Kishimoto1999}, and allowed accurate determination of the 
position of the source of scattered radiation \citep{Capetti1995}. Heavily obscured AGN (such as Mrk~231) were observed to probe 
the composition of dust and low-ionization gas clouds \citep{Smith1995}. UV polarization also helped unveil the characteristics 
of the magnetic-field pattern in the jet of M87 \citep{Boksenberg1992} and probed the synchrotron origin of optical polarization 
in the BL Lac object PKS~2155-304 \citep{Allen1993}.

Both WUPPE and HST polarimetric observations brought important results in the field of AGN. They were, however, restricted to 
low-resolution capabilities (FOS linewidths 1.89 -- 1.97~\AA, for a 3.7" x 1.3" and 0.26" aperture, respectively) and did not reach
wavelengths below 1150~\AA. This is unfortunate, because polarization induced by scattering on small dust grains rises steeply 
into the blue (1200 -- 3600~\AA, \citealt{Kartje1995}). Moreover, contamination by the background starlight of AGN-host galaxies 
is about three orders of magnitude lower at 0.1~$\mu$m than at 1~$\mu$m (for spiral galaxies, see \citealt{Bolzonella2000}). Hence, 
the contrast of polarimetric observations is expected to increase significantly from longer to shorter wavelengths, leaving today
a vast new parameter space to be explored by a new high-resolution instrument. 

\section{The LUVOIR mission and the POLLUX instrument}
The Large Ultraviolet/Optical/Infrared Surveyor (LUVOIR) is one of four ``flagship'' mission concept studies led by NASA for 
the 2020 Decadal Survey. LUVOIR is a concept for an ambitious, multi-wavelengths 15-m observatory that would enable a 
great leap forward in a broad range of astrophysical topics, from the epoch of re-ionization, through galaxy formation and evolution, 
to star and planet formation. LUVOIR also has the major goal of characterizing a wide range of exoplanets, including those 
that might be habitable - or even inhabited. If LUVOIR is selected during the Decadal evaluation, this mission would be launched
in 2035. 

The study of LUVOIR will extend over three years and be executed by the Goddard Space Flight Center, under the leadership
of a Science and Technology Definition Team (STDT). Under the impulsion of the Laboratoire d'Astrophysique de Marseille (LAM) 
and the Laboratoire d'\'etudes spatiales et d'instrumentation en astrophysique (LESIA), European institutes have come 
together to propose an instrument that would be on-board the 15-meter primary mirror option of LUVOIR. This instrument, 
POLLUX, is a high-resolution spectropolarimeter operating at UV wavelengths (900 -- 4000~\AA). LUVOIR will be equipped with 4 instruments:
1) a coronagraph called ECLIPS, 2) HDI, a near-UV to near-IR imager, 3) a multi-object low and medium resolution UV 
spectrograph and imager called LUMOS, and 4) POLLUX. The first 3 instruments are being studied by NASA, while POLLUX is 
being studied by a European consortium led by France. 

In its actual design, POLLUX would resolve narrow UV emission and absorption lines, following the various forms of AGN feedback 
into the interstellar and intergalactic medium. The most innovative characteristic of POLLUX is its unique spectropolarimetric
capability that will enable detection of the UV circular and linear polarization from almost all types of sources, providing a full 
picture of their scattering and magnetic field properties. Since the parameter space opened by POLLUX is essentially uncharted territory,
its potential for ground-breaking discoveries is tremendous. It will also neatly complement and enrich some of the cases advanced for
LUMOS, the multi-object spectrograph of LUVOIR.

\section{AGN science with POLLUX}
POLLUX will offer unique insights into the still poorly-known physics of AGN, in particular by probing
UV-emitting and absorbing material arising from accretion disks, synchrotron emission in jet-dominated AGN and 
large-scale outflows. Some key signatures of accretion disks can be revealed only in polarized light, and with 
higher contrast at ultraviolet than at longer wavelengths. Specifically, models of disk atmospheres usually assume 
Compton scattering in an electron-filled plasma, resulting in inclination-dependent polarization signatures 
(up to 10\%, see e.g., \citealt{Chandrasekhar1960}). Yet optical polarization is detected at less than a percent, and 
parallel to the radio jets if any \citep{Stockman1979}. Whether these low levels can be attributed to dominant 
absorption opacity \citep{Laor1989} or complete Faraday depolarization \citep{Agol1996} is unclear. This degeneracy 
can be broken by looking at the numerous UV spectral lines that are formed in the innermost AGN regions (e.g, 
Ly$\alpha$~$\lambda$1216, C~{\sc ii}~$\lambda$1335, C~{\sc iv}~$\lambda$1549, Mg~{\sc ii}~$\lambda$2800 ...). These 
lines are the key to understanding UV polarization, and only observations with high signal-to-noise ratio and
high spectral resolution can distinguish between the two effects. If absorption opacity is responsible for the 
low continuum polarization we detect, the line profiles should also show a significant drop in polarization.

Another interesting feature coupled with the accretion disk is the strong polar magnetic field that ultimately 
launches jets. Dissipative processes in the accretion disk transfer matter inward, angular momentum outward, 
and heat up the disk. Magnetic-field lines from the inner part of the accretion disk cross the event horizon of 
the black hole and are wound up by its spin, launching Poynting flux-dominated outflows. The resulting jets tend
to be collimated for a few parsecs and to dilute in giant lobes on kilo-parsec scales. Relativistic electrons 
traveling in ordered magnetic fields are responsible for the high polarization we detect (of the order of 40 -- 60\%, 
see e.g., \citealt{Thomson1995}). Interestingly, the continuum-polarization degree and angle are extremely sensitive 
to the strength and direction of the magnetic field, and to the charge distribution. This will allow POLLUX to probe 
in great detail the magnetic configuration of such jets by measuring the electron-beam polarization. If a jet is 
inclined toward the observer (blazar-like objects), a non-thermal spectral energy distribution will be observed, 
with a low-energy broadband peak in the radio-to-UV wavelength range. Comparing the observed UV polarization of 
blazars to leptonic, hadronic or alternative jet models (e.g., \citealt{Zhang2017}) will enable better constraints 
on the composition and lifetimes of particles in the plasma. Since jets are also responsible for ion and neutrino 
emission, they are valuable sources to understand how cosmic rays are produced. 

In addition to jets, strong polar outflows will be important targets for POLLUX. At redshift greater than 1.5 -- 2, 
a sub-category of quasars, called Broad-Absorption-Line quasars (BAL QSO), exhibit very broad absorption 
features in UV resonant lines (Ly$\alpha$, C~{\sc iv}, Si~{\sc iv}). The gas outflows producing these signatures
presumably contribute to the enrichment of the quasar host galaxies (a process generally referred to as `feedback'). BAL QSO are 
particularly interesting as they tend to have high polarization degrees \citep[e.g.,][]{Ogle1999}, which can be used to 
constrain wind geometry \citep{Young2007}. These BAL QSO are believed to be the high-redshift analogs 
of more nearby, polar-scattered Seyfert galaxies, whose UV emission will also be explorable with POLLUX. In particular, 
POLLUX will help investigate the dependence of broad absorption lines on bolometric luminosity and thus 
the role of radiative acceleration in the appearance of these lines \citep{Arav1994a,Arav1994b}. High-resolution 
spectropolarimetry will also enable new constraints on wind kinematics, for the first time from UV resonance lines,
similarly to what has been achieved by \citet{Young2007} using the H$\alpha$ line.

Combined with the UV, optical and IR capabilities of the other instruments on board LUVOIR, POLLUX 
will allow unprecedented insight into the composition of AGN dust. In the Galaxy, the dust extinction,
which is highest in the UV, shows a local peak near 2175~\AA\ \citep{Stecher1965}. 
The strength of this feature varies from galaxy to galaxy: it is weaker in the Large and Small Magellanic
Clouds than in the Galaxy and almost never observed in AGN \citep{Gaskell2007}. 
Unveiling the mineralogy of extragalactic dust grains is not easy and requires high-quality extinction-curve 
measurements. A strong advantage for POLLUX is that the polarization induced by dust scattering rises rapidly 
toward the blue, peaking near 3000~\AA~in the rest frame and remaining nearly constant at shorter wavelengths 
(see, e.g., \citealt{Hines2001}). Polarimetry at short wavelengths can thus discriminate between dust scattering and 
wavelength-independent electron scattering. Additionally, polarization measurements with POLLUX can be enhanced 
by dust-grain alignment: theory predicts that paramagnetic grains will be aligned with their longer axes 
perpendicular to the local magnetic field if exposed to magnetic or anisotropic-radiation fields with wavelengths
less than the grain diameter \citep{Lazarian2007}. Therefore, 
the UV band will selectively trace the smallest dust grains and allow better characterization of AGN dust 
composition. For such grains, the polarization strength is predicted to be proportional to the magnetic-field 
strength, enabling POLLUX to also measure for the first time the intensity and direction of the magnetic field 
on parsec scales around the AGN core \citep{Hoang2014}. Finally, the radiative pumping of atoms and ions 
with fine structure is predicted to align these with the magnetic field, giving rise to polarized-line emission. 
A number of prominent UV lines are predicted to show significant polarization following that mechanism, 
providing a mean of tracing the magnetic field in hot AGN gas on small scales \citep{Yan2008}.

\section{Conclusions}
We have highlighted the need for a high-resolution spectropolarimeter covering the full ultraviolet band to study 
a wealthy range of astrophysical sources, emphasizing on the still poorly-constrained AGN physics. Concretely, POLLUX 
will probe the location, geometry and composition of the regions responsible for UV emission and enable measurements
of their magnetic-field strength and topology. Measurements of UV polarization due to dust scattering by magnetically 
aligned grains will allow one to assess the strength and direction of the magnetic fields that are shaping the AGN 
outskirts. Outflows, jets and feedback, which drive the co-evolution of the AGN and their host galaxies, will be 
probed at unprecedented resolution.

The AGN science case of POLLUX would fully benefit from the broad wavelength coverage of LUVOIR to investigate 
the properties of accretion disks and jets. By the time LUVOIR would be launched ($\ge$ 2035), it would fully 
take advantage of X-ray coverage by ATHENA observations to probe the physics of accretion \citep{Nandra2013}. 
From the ground, sub-millimeter, millimeter and radio observations from large arrays of antennae such as SKA 
\citep{Acero2017} will probe the low-energy end of AGN spectra, together with high-resolution images 
of the central parsecs and jets. Interferometry, in particular in the infrared domain, will enable subparsec-resolution 
images of the hot and cold dust components.

\begin{acknowledgements}
FM thanks the Centre national d'\'etudes spatiales (CNES) who funded his research through the post-doctoral grant 
``Probing the geometry and physics of active galactic nuclei with ultraviolet and X-ray polarized radiative transfer''. 
\end{acknowledgements}

\bibliographystyle{aa}  
\bibliography{Marin} 

\begin{thebibliography}{30}
\expandafter\ifx\csname natexlab\endcsname\relax\def\natexlab#1{#1}\fi

\bibitem[{{Acero} {et~al.}(2017){Acero}, {Acquaviva}, {Adam}, {Aghanim},
  {Allen}, {Alves}, {Ammanouil}, {Ansari}, {Araudo}, {Armengaud}, {Ascaso},
  {Athanassoula}, {Aubert}, {Babak}, {Bacmann}, {Banday}, {Barriere},
  {Bellossi}, {Bernard}, {Bernardini}, {B{\'e}thermin}, {Blanc}, {Blanchet},
  {Bobin}, {Boissier}, {Boisson}, {Boselli}, {Bosma}, {Bosse}, {Bottinelli},
  {Boulanger}, {Boyer}, {Bracco}, {Briand}, {Bucher}, {Buat}, {Cambresy},
  {Caillat}, {Casandjian}, {Caux}, {C{\'e}lestin}, {Cerruti}, {Charlot},
  {Chassande-Mottin}, {Chaty}, {Christensen}, {Ciesla}, {Clerc},
  {Cohen-Tanugi}, {Cognard}, {Combes}, {Comis}, {Corbel}, {Cordier}, {Coriat},
  {Courtin}, {Courtois}, {Da Silva}, {Daddi}, {Dallier}, {Dartois}, {Demyk},
  {Denis}, {Denis}, {Djannati-Ata{\"i}}, {Donati}, {Douspis}, {van Driel}, {El
  Korso}, {Falgarone}, {Fantina}, {Farges}, {Ferrari}, {Ferrari},
  {Ferri{\`e}re}, {Flamary}, {Gac}, {Gauffre}, {Genova}, {Girard}, {Grenier},
  {Griessmeier}, {Guillard}, {Guillemot}, {Gulminelli}, {Gusdorf}, {Habart},
  {Hammer}, {Hennebelle}, {Herpin}, {Hervet}, {Hughes}, {Ilbert}, {Janvier},
  {Josselin}, {Julier}, {Lachaud}, {Lagache}, {Lallement}, {Lambert}, {Lamy},
  {Langer}, {Larzabal}, {Lavaux}, {Le Bertre}, {Le F{\`e}vre}, {Le Tiec},
  {Lefloch}, {Lehnert}, {Lemoine-Goumard}, {Levrier}, {Limousin}, {Lis},
  {L{\'o}pez-Sepulcre}, {Macias-Perez}, {Magneville}, {Marcowith}, {Margueron},
  {Marquette}, {Marshall}, {Martin}, {Mary}, {Masson}, {Maurogordato},
  {Mazauric}, {Mellier}, {Miville-Desch{\^e}nes}, {Montier}, {Mottez},
  {Mourard}, {Nesvadba}, {Nezan}, {Noterdaeme}, {Novak}, {Ocvirk}, {Oertel},
  {Olive}, {Ollier}, {Palanque-Delabrouille}, {Pandey-Pommier}, {Pennec},
  {P{\'e}rault}, {Peroux}, {Petit}, {P{\'e}tri}, {Petiteau}, {Pety}, {Pratt},
  {Puech}, {Quertier}, {Raffin}, {Rakotozafy Harison}, {Rawson}, {Renaud},
  {Revenu}, {Richard}, {Richard}, {Rincon}, {Ristorcelli}, {Rodriguez},
  {Schultheis}, {Schimd}, {Semelin}, {Sol}, {Starck}, {Tagger}, {Tasse},
  {Theureau}, {Torchinsky}, {Vastel}, {Vergani}, {Verstraete}, {Vigouroux},
  {Vilmer}, {Vilotte}, {Webb}, {Ysard}, \& {Zarka}}]{Acero2017}
{Acero}, F., {Acquaviva}, J.-T., {Adam}, R., {et~al.} 2017, ArXiv e-prints

\bibitem[{{Agol} \& {Blaes}(1996)}]{Agol1996}
{Agol}, E. \& {Blaes}, O. 1996, \mnras, 282, 965

\bibitem[{{Allen} {et~al.}(1993){Allen}, {Smith}, {Angel}, {Miller},
  {Anderson}, \& {Margon}}]{Allen1993}
{Allen}, R.~G., {Smith}, P.~S., {Angel}, J.~R.~P., {et~al.} 1993, \apj, 403,
  610

\bibitem[{{Antonucci} {et~al.}(1994){Antonucci}, {Hurt}, \&
  {Miller}}]{Antonucci1994}
{Antonucci}, R., {Hurt}, T., \& {Miller}, J. 1994, \apj, 430, 210

\bibitem[{{Arav} \& {Li}(1994)}]{Arav1994a}
{Arav}, N. \& {Li}, Z.-Y. 1994, \apj, 427, 700

\bibitem[{{Arav} {et~al.}(1994){Arav}, {Li}, \& {Begelman}}]{Arav1994b}
{Arav}, N., {Li}, Z.-Y., \& {Begelman}, M.~C. 1994, \apj, 432, 62

\bibitem[{{Boksenberg} {et~al.}(1992){Boksenberg}, {Macchetto}, {Albrecht},
  {Barbieri}, {Blades}, {Crane}, {Deharveng}, {Disney}, {Jakobsen},
  {Kamperman}, {King}, {Mackay}, {Paresce}, {Weigelt}, {Baxter}, {Greenfield},
  {Jedrzjewski}, {Nota}, \& {Sparks}}]{Boksenberg1992}
{Boksenberg}, A., {Macchetto}, F., {Albrecht}, R., {et~al.} 1992, \aap, 261,
  393

\bibitem[{{Bolzonella} {et~al.}(2000){Bolzonella}, {Miralles}, \&
  {Pell{\'o}}}]{Bolzonella2000}
{Bolzonella}, M., {Miralles}, J.-M., \& {Pell{\'o}}, R. 2000, \aap, 363, 476

\bibitem[{{Capetti} {et~al.}(1995){Capetti}, {Macchetto}, {Axon}, {Sparks}, \&
  {Boksenberg}}]{Capetti1995}
{Capetti}, A., {Macchetto}, F., {Axon}, D.~J., {Sparks}, W.~B., \&
  {Boksenberg}, A. 1995, \apjl, 452, L87

\bibitem[{{Chandrasekhar}(1960)}]{Chandrasekhar1960}
{Chandrasekhar}, S. 1960, {Radiative transfer}

\bibitem[{{Code} {et~al.}(1993){Code}, {Meade}, {Anderson}, {Nordsieck},
  {Clayton}, {Whitney}, {Magalhaes}, {Babler}, {Bjorkman}, {Schulte-Ladbeck},
  \& {Taylor}}]{Code1993}
{Code}, A.~D., {Meade}, M.~R., {Anderson}, C.~M., {et~al.} 1993, \apjl, 403,
  L63

\bibitem[{{Code} \& {Nordsieck}(1989)}]{Code1989}
{Code}, A.~D. \& {Nordsieck}, K.~H. 1989, in \baas, Vol.~21, Bulletin of the
  American Astronomical Society, 756

\bibitem[{{Gaskell} \& {Benker}(2007)}]{Gaskell2007}
{Gaskell}, C.~M. \& {Benker}, A.~J. 2007, ArXiv e-prints

\bibitem[{{Hines} {et~al.}(2001){Hines}, {Schmidt}, {Gordon}, {Smith}, {Wills},
  {Allen}, \& {Sitko}}]{Hines2001}
{Hines}, D.~C., {Schmidt}, G.~D., {Gordon}, K.~D., {et~al.} 2001, \apj, 563,
  512

\bibitem[{{Hoang} {et~al.}(2014){Hoang}, {Lazarian}, \& {Martin}}]{Hoang2014}
{Hoang}, T., {Lazarian}, A., \& {Martin}, P.~G. 2014, \apj, 790, 6

\bibitem[{{Kartje}(1995)}]{Kartje1995}
{Kartje}, J.~F. 1995, \apj, 452, 565

\bibitem[{{Kishimoto}(1999)}]{Kishimoto1999}
{Kishimoto}, M. 1999, \apj, 518, 676

\bibitem[{{Laor} \& {Netzer}(1989)}]{Laor1989}
{Laor}, A. \& {Netzer}, H. 1989, \mnras, 238, 897

\bibitem[{{Lazarian} \& {Hoang}(2007)}]{Lazarian2007}
{Lazarian}, A. \& {Hoang}, T. 2007, \apjl, 669, L77

\bibitem[{{Nandra} {et~al.}(2013){Nandra}, {Barret}, {Barcons}, {Fabian}, {den
  Herder}, {Piro}, {Watson}, {Adami}, {Aird}, {Afonso}, \& et~al.}]{Nandra2013}
{Nandra}, K., {Barret}, D., {Barcons}, X., {et~al.} 2013, ArXiv e-prints

\bibitem[{{Nordsieck} \& {Code}(1982)}]{Nordsieck1982}
{Nordsieck}, K.~H. \& {Code}, A.~D. 1982, in \baas, Vol.~14, Bulletin of the
  American Astronomical Society, 657

\bibitem[{{Ogle} {et~al.}(1999){Ogle}, {Cohen}, {Miller}, {Tran}, {Goodrich},
  \& {Martel}}]{Ogle1999}
{Ogle}, P.~M., {Cohen}, M.~H., {Miller}, J.~S., {et~al.} 1999, \apjs, 125, 1

\bibitem[{{Smith} {et~al.}(1995){Smith}, {Schmidt}, {Allen}, \&
  {Angel}}]{Smith1995}
{Smith}, P.~S., {Schmidt}, G.~D., {Allen}, R.~G., \& {Angel}, J.~R.~P. 1995,
  \apj, 444, 146

\bibitem[{{Stanford} {et~al.}(1985){Stanford}, {Murison}, {Whitney}, \&
  {Clayton}}]{Stanford1985}
{Stanford}, S.~A., {Murison}, M.~A., {Whitney}, B.~A., \& {Clayton}, G.~C.
  1985, in \baas, Vol.~17, Bulletin of the American Astronomical Society, 900

\bibitem[{{Stecher} \& {Donn}(1965)}]{Stecher1965}
{Stecher}, T.~P. \& {Donn}, B. 1965, \apj, 142, 1681

\bibitem[{{Stockman} {et~al.}(1979){Stockman}, {Angel}, \&
  {Miley}}]{Stockman1979}
{Stockman}, H.~S., {Angel}, J.~R.~P., \& {Miley}, G.~K. 1979, \apjl, 227, L55

\bibitem[{{Thomson} {et~al.}(1995){Thomson}, {Robinson}, {Tanvir}, {Mackay}, \&
  {Boksenberg}}]{Thomson1995}
{Thomson}, R.~C., {Robinson}, D.~R.~T., {Tanvir}, N.~R., {Mackay}, C.~D., \&
  {Boksenberg}, A. 1995, \mnras, 275, 921

\bibitem[{{Yan} \& {Lazarian}(2008)}]{Yan2008}
{Yan}, H. \& {Lazarian}, A. 2008, \apj, 677, 1401

\bibitem[{{Young} {et~al.}(2007){Young}, {Axon}, {Robinson}, {Hough}, \&
  {Smith}}]{Young2007}
{Young}, S., {Axon}, D.~J., {Robinson}, A., {Hough}, J.~H., \& {Smith}, J.~E.
  2007, \nat, 450, 74

\bibitem[{{Zhang}(2017)}]{Zhang2017}
{Zhang}, H. 2017, Galaxies, 5, 32

\end{thebibliography}

\end{document}